\documentclass[aps,prb,twocolumn,showpacs,notitlepage,10pt,superscriptaddress]{revtex4-1}

\pdfoutput=1
\usepackage{times}
\usepackage{epsfig}
\usepackage{amsmath}
\usepackage{amssymb}
\usepackage{wasysym}
\usepackage{MnSymbol}
\usepackage[colorlinks=true,citecolor=blue,linkcolor=red]{hyperref}
\newcommand{\bs}{\boldsymbol}

\begin{document}

\title{Corner states of topological fullerenes}
\author{Andreas R\"uegg}
\affiliation{Department of Physics, University of California,
  Berkeley, Berkeley CA 94720, USA}
  \affiliation{Theoretische Physik, ETH Z\"urich, CH-8093 Z\"urich, Switzerland}
\author{Sinisa Coh}
\affiliation{Department of Physics, University of California,
  Berkeley, Berkeley CA 94720, USA}
\affiliation{Materials Science Division, Lawrence Berkeley National
  Laboratory, Berkeley CA 94720, USA}
\author{Joel E.~Moore}
\affiliation{Department of Physics, University of California,
  Berkeley, Berkeley CA 94720, USA}
\affiliation{Materials Science Division, Lawrence Berkeley National
  Laboratory, Berkeley CA 94720, USA}
\date{\today}
\begin{abstract}
  The unusual electronic properties of the quantum spin Hall or Chern
  insulator become manifest in the form of robust edge states when
  samples with boundaries are studied. In this work, we ask if and how
  the topologically non-trivial electronic structure of these
  two-dimensional systems can be passed on to their zero-dimensional
  relatives, namely fullerenes or other closed-cage molecules. To
  address this question, we study Haldane's honeycomb lattice model
  on polyhedral nano-surfaces.  We find that for sufficiently large surfaces 
  characteristic corner states appear for parameters for which the planar model
  displays a quantized Hall effect.
  In the electronic structure, these corner states show up as in-gap
  modes which are well separated from the quasi-continuum of
  states. We discuss the role of finite size effects and how the
  coupling between the corner states lifts the degeneracy in a 
  characteristic way determined by the combined Berry phases which leads to an 
  effective magnetic monopole of charge 2 at the center of the nano-surface. 
  Experimental implications for fullerenes in the 
  large spin-orbit regime are also pointed out.
\end{abstract}
\maketitle
\section{Introduction}

\begin{figure*}
\includegraphics[width=1.9\columnwidth]{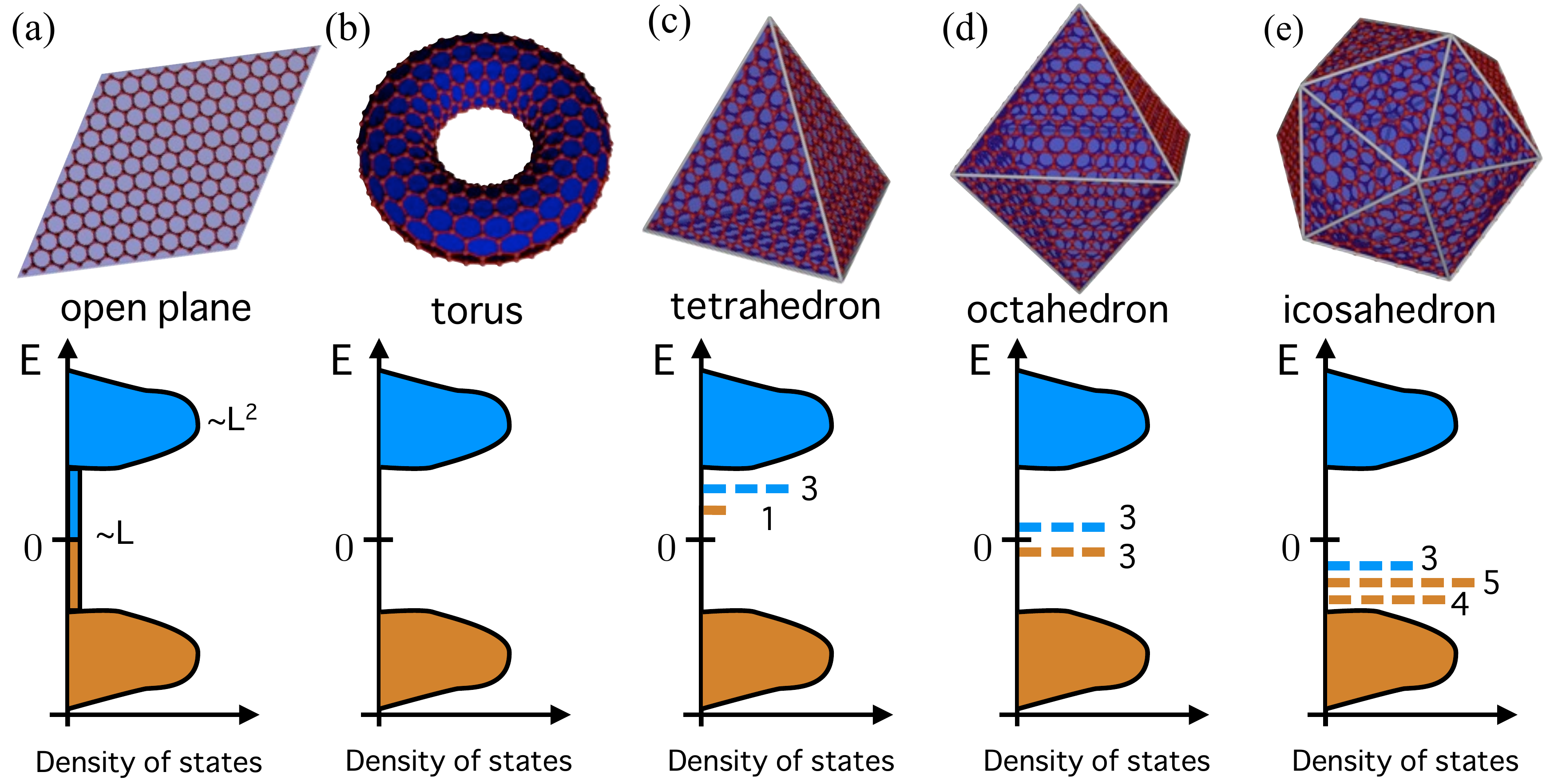}
\caption{(Color online.) Schematics of the electronic spectrum of the
  Haldane model in the Chern insulator phase on various
  geometries. (a) For a finite open system, edge states appear in the
  gap. The number of in-gap states is proportional to the
  circumference $L$ of the sample while the number of states in the
  valence or conduction bands is proportional to $L^2$. (b) If
  periodic boundary conditions are employed to form a torus, the
  spectrum is gapped as for the infinite planar model. If the Haldane
  model is studied on polyhedral surfaces (c), (d) and (e), a finite
  number of in-gap states is observed. The number of in-gap states
  depends on the geometry of the nano-surface, namely the number of
  corners. Moreover, the degeneracy of the in-gap levels is lifted in
  a characteristic way as indicated. In the bottom panels, occupied
  states are colored orange, while empty states are in blue.}
\label{fig:polyhedra}
\end{figure*}

The non-trivial topological electronic properties of two-dimensional
Chern insulators\cite{Haldane:1988,Chang:2013} (quantum anomalous Hall
insulators) or quantum spin Hall
insulators\cite{Kane:2005a,Kane:2005b,Bernevig:2006,
  Bernevig:2006a,Konig:2007} imply the existence of topologically
protected boundary modes in systems with boundaries. While the chiral
edge states of the Chern insulator are immune to backscattering and
hence robust against all forms of weak disorder,\cite{Haldane:1995}
the helical edge states of a quantum spin Hall insulator are at least
protected against elastic scattering off non-magnetic
impurities until the edge electron-electron interactions are rather strong.\cite{Xu:2006,Wu:2006}
In both cases, unless disorder is so strong
as to drive a phase transition, edge states are present independent of the shape or
microscopic structure of the boundary. Because these boundary modes
live in the bulk gap of the single-particle spectrum, they appear as
in-gap levels in the total density-of-states, see
Fig.~\ref{fig:polyhedra}~(a). This provides a way to distinguish a
topological from a trivial insulator for which edge states are
generically absent and the spectrum remains gapped in the presence of
a boundary.
On the other hand, if we use periodic instead of open boundary
conditions [i.e.~consider the system on a torus,
Fig.~\ref{fig:polyhedra}~(b)], edge states are gapped out. In this
case, the spectrum of a topological insulator is indistinguishable
from the spectrum of a trivial insulator. One might expect that this
conclusion remains valid if the system is put on any closed (meaning,
without boundary) surface.

In this article, by studying Haldane's honeycomb lattice
model\cite{Haldane:1988} on topologically {\it spherical} nano-surfaces
(i.e.~polyhedra), we provide counterexamples to this naive
expectation. Namely, we demonstrate that on such closed but
sufficiently large surfaces, the non-trivial topological invariant of
the two-dimensional model is revealed in the electronic spectrum:
choosing parameters for which the planar system has a non-vanishing
Chern number $C=\pm 1$, we identify characteristic in-gap states which
are well separated from the quasi-continuum of the remaining levels.
Moreover, we find that these in-gap levels correspond to eigenstates
which are localized at the {\it corners} of the polyhedral surfaces. 
In analogy with the closed-cage molecules formed from carbon
atoms,\cite{Dresselhaus:1996Book} we dub the systems displaying the
characteristic corner states as {\it topological fullerenes}. A
summary of our main results is illustrated in
Fig.~\ref{fig:polyhedra}.  To avoid confusion, we stress that our
nomenclature does {\it not} refer to a topological invariant of a
zero-dimensional free fermion system.\cite{Kitaev:2009,Ryu:2010b} [An
example of such a zero-dimensional invariant was given by
Kitaev:\cite{Kitaev:2009} in the absence of time-reversal symmetry
(class A), the number of occupied states below the Fermi energy
determines a $\mathbb{Z}$ index.]
Instead, we ask the question of what happens to a two-dimensional
topologically non-trivial system if it is put on a closed
two-dimensional nano-surface. Hence, the name ``topological
fullerenes'' solely refers to the topologically non-trivial properties
of the two-dimensional parent system.

In passing, we note that closed (spherical) surfaces with quantized
Hall conductivity similar to the ones studied in this paper also
appear when the orbital magneto-electric
effect is analyzed via the theory of electrical polarization:\cite{Essin:2009} for a 3D solid, the orbital-electronic
contribution to the magneto-electric coupling has a quantum of
indeterminacy. This quantum corresponds to the possibility of
absorbing layers with quantized Hall conductivity on the surfaces of
the solid.

Our theoretical analysis focuses on the tetrahedral, octahedral and
icosahedral nano-surfaces. These polyhedral surfaces can be
constructed from the planar honeycomb lattice by cutting out
appropriate wedges and gluing the edges back
together.\cite{Gonzalez:1993} While spherical carbon
fullerenes,\cite{Dresselhaus:1996Book,Kolesnikov:2009} such as the
$C_{60}$ buckyball, have the shape of an icosaherdon, materials like
boron nitride\cite{Goldberg:1998} or transition-metal
dichalcogenides\cite{Parilla:2004,Wang:2012fk} prefer to form octahedral
fullerenes. We are not aware of a material which realizes a
tetrahedral nano-surface but from a theoretical perspective it is
instructive (and simple enough) to include this surface in our
discussion as well.

To date, we do not know of an experimental system realizing Haldane's
honeycomb lattice model. However, there are interesting proposals that
a time-reversal invariant quantum spin Hall insulator with non-trivial
$\mathbb{Z}_2$ index can be realized on the honeycomb
lattice.\cite{Kane:2005a,Kane:2005b} The first route to stabilizing
such a phase considers the possibility of inducing a large spin-orbit
coupling in graphene via heavy adatoms.\cite{Weeks:2011,Hu:2012} The
second approach seeks for alternative graphene-like materials with
large intrinsic spin-orbit coupling, such as a single
Bi-bilayer,\cite{Murakami:2006} silicene (2D-Si)\cite{LiuC:2011} or
2D-tin.\cite{Xu:2013} There are first experimental indications for the
existence of a topological insulator phase in
Bi-bilayers\cite{Sabater:2013} and it is reasonable to assume that if
the 2D versions exists, also closed-cage molecules might be synthesized,
eventually.

The remainder of the paper is organized as follows: In
Sec.~\ref{sec:tb} we relate the corners of the polyhedral surfaces to
topological lattice defects called {\it disclinations} and we specify
how to define Haldane's model on the considered nano-surfaces. In
Sec.~\ref{sec:disclination} we briefly review the properties of an
isolated disclination in the Haldane model and provide a topological
perspective on the existence of non-trivial bound states. In
Sec.~\ref{sec:num} we present results from numerically diagonalizing
various polyhedral systems to demonstrate the existence of the corner
states. We also discuss the finite size effects which should be small
in order to identify the in-gap states. In Sec.~\ref{sec:eff}, we
investigate how the degeneracy of the in-gap levels is lifted due to
the coupling between the corner states in a finite system. To recover the observed splitting, we include Berry phase terms which can be represented as an effective magnetic monopole of charge 2 at the center of the polyhedral surfaces. We conclude
in Sec.~\ref{sec:conclusion} by summarizing our results and providing
a more detailed discussion of possible experimental systems.
\section{Model for topological fullerenes}
\label{sec:tb}
\subsection{Polyhedral nano-surfaces}
To study topological effects on closed-cage molecules, we first
generalize Haldane's Chern insulator model on the honeycomb lattice to
polyhedral nano-surfaces. It is well-known that a polyhedral
nano-surface cannot be formed using only
hexagons.\cite{Dresselhaus:1996Book} Instead, $n$-gons with $n<6$ have
to be introduced and in the following we briefly discuss the general
structure of such molecules. The fundamental relation satisfied by all
the closed nano-surfaces is given by Euler's famous formula
\begin{equation}
V-E+F=\chi.
\label{eq:Euler}
\end{equation}
Equation~\eqref{eq:Euler} relates the number of faces $F$, the number
of vertices $V$ and the number of edges $E$ to the Euler
characteristic $\chi$. For a spherical polyhedral surface, $\chi=2$
while for the torus $\chi=0$. For a given $n<6$, one can now easily
compute the number $N$ of $n$-gons which are required in addition to
the number $H$ of hexagons to form a closed surface, by noting that
\begin{eqnarray*}
F&=&N+H\\
2E&=&nN+6H\\
3V&=&nN+6H.
\end{eqnarray*}
In combination with Eq.~\eqref{eq:Euler}, $N$ can be obtained
\begin{equation}
N=\frac{\chi}{1-n/6}=\frac{6\chi}{f},
\end{equation}
where $f=6-n$. [Note that $H$ is undetermined by
Eq.~\eqref{eq:Euler}.] For the torus ($\chi=0$), it follows that $N=0$
and no defects are necessary.\footnote{An even number of defects with
  opposite $f$ is also possible on the torus, for example two pentagon
  and two heptagon defects.}  On the other hand, for the polyhedral
surfaces ($\chi=2$), $N$ is non-vanishing. Specifically, an
icosahedral surface can be formed with additional 12 pentagons
($f=1$), an octahedral surface with additional 6 squares ($f=2$) or a
tetrahedral surface with additional 4 triangles ($f=3$). In essence,
the total curvature needed to form a sphere-like molecule with
$\chi=2$ is concentrated at the $n$-gons with $n<6$. Hence, the
$n$-gonal lattice defects form the corners of the polyhedra and, as
discussed below, are crucial for understanding the electronic
structure of the Haldane model defined on these nanosurfaces.

\subsection{Tight-binding model}
We are now in a position to define the Haldane model on the
polyhedral surfaces discussed above. The tight-binding model is given
by\cite{Haldane:1988}
\begin{equation}
  \mathcal{H}=-t\sum_{\langle i,j\rangle}(c_i^{\dag}c_j+{\rm
    h.c.})-t_2\sum_{\langle\!\langle
    i,j\rangle\!\rangle}(e^{-i\phi_{ij}}c_i^{\dag}c_j+{\rm
    h.c.})+\mathcal{H}_{\Delta}.
\label{eq:Haldane}
\end{equation}
Here, $c_i^{\dag}$ and $c_i$ are fermionic creation and annihilation
operators of spinless electrons on site $i$, respectively. The
nearest-neighbor hopping amplitude is $t=1$ which sets the unit of
energy and $t_2$ is the second-neighbor hopping with phase factors
$e^{i\phi_{ij}}$. In the planar model, the phases $\phi_{ij}$ are
chosen such that a staggered flux configuration, which preserves both
the original unit cell and the six-fold rotation symmetry, is
realized.\cite{Haldane:1988} For the studied nano-surfaces, we use the
bulk assignment of $\phi_{ij}$ for all the hexagons. Indeed, it is
possible to choose the handedness of $\phi_{ij}$ consistently on all
the faces and across the edges where they meet. Using the concept of a
local Chern vector as introduced recently in
Ref.~\onlinecite{Bianco:2011}, this choice guarantees a local Chern
vector which always points either outward or inward of the surface.
For the most part, we will set $\phi_{ij}=\pm\pi/2$ such that the
second-neighbor hopping is purely imaginary. Across the $n$-gons with
$n=3$, 4 or 5, the phase factors are not uniquely defined and we
therefore choose $t_2=0$. However, the results to be derived do not depend
in an important way on the choice of the second-neighbor hopping
across the $n$-gons, as they can be obtained from general analytical
arguments that are independent of this choice.

The last term $\mathcal{H}_{\Delta}$ in Eq.~\eqref{eq:Haldane} is
identical to zero for the tetrahedral and icosahedral nano-surfaces,
$\mathcal{H}_{\Delta}=0$. For the octahedral surfaces, on the other
hand, it is defined as
\begin{equation}
  \mathcal{H}_{\Delta}=\Delta\left(\sum_{i\in A} n_i-\sum_{i\in
      B}n_i\right),
\label{eq:HDelta}
\end{equation}
where $n_i=c_i^{\dag}c_i$ and $A$ and $B$ refer to the two
sublattices. The staggered sublattice potential $\mathcal{H}_{\Delta}$
can stabilize a topologically trivial phase in the planar
system.\cite{Haldane:1988} The definition Eq.~\eqref{eq:HDelta}
requires a {\it global} assignment of two sublattices $A$ and $B$. We
therefore include the staggered sublattice potential only on the
octahedral surfaces. Both tetrahedral and icosahedral surfaces do not
allow for a global definition of two sublattices and attempting to
define Eq.~\eqref{eq:HDelta} would require to introduce domain walls
across which the definition of the A-B sublattices changes.
\section{Isolated disclination}
\label{sec:disclination}
\subsection{Overview of results}
\begin{figure}
\includegraphics[width=0.9\linewidth]{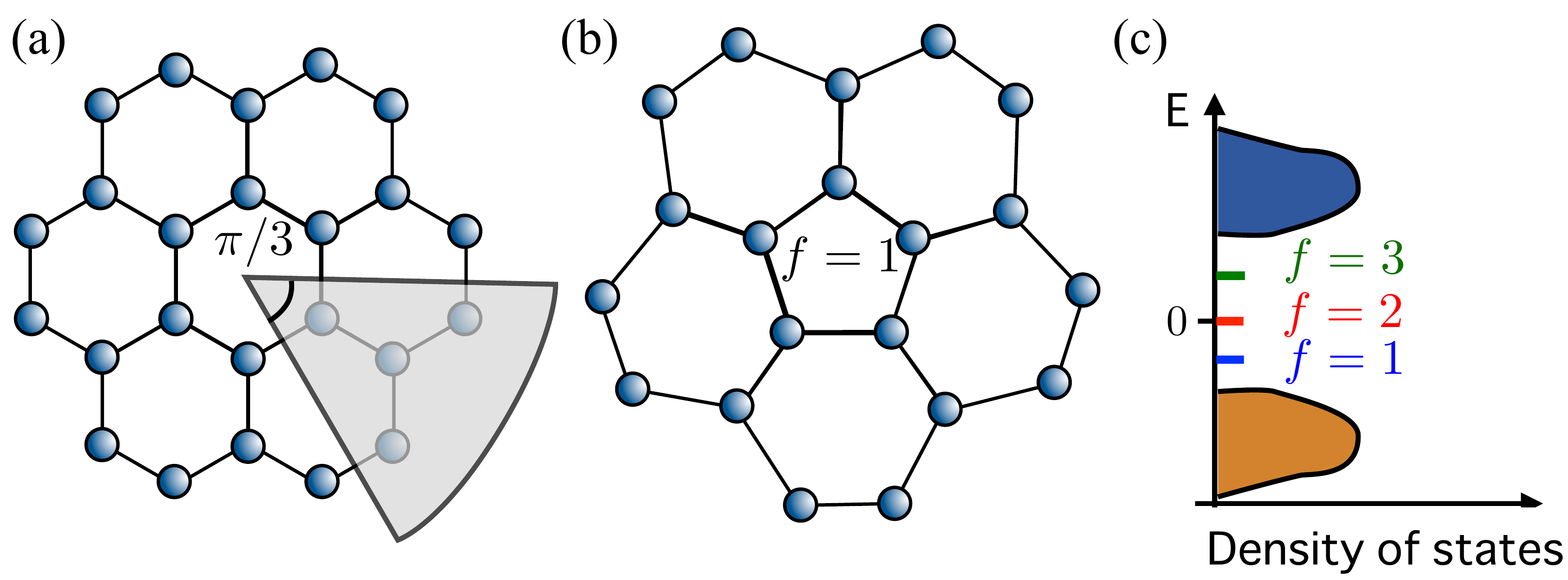}
\caption{(Color online.) (a) An isolated wedge disclination is
  constructed by cutting out $f$ times a $\pi/3$-wedge and gluing the
  two sides back together. (b) For $f=1$, the disclination core is
  formed by a pentagon. (c) In the Haldane model, different types of
  disclinations induce in-gap states with different
  energies.}
\label{fig:disclination}
\end{figure}

The $n$-gonal lattice defects appearing at the corners of the
polyhedral surfaces are known as wedge
disclinations\cite{Chaikin:2000} and are characterized by the {\it
  Frank index} $f=6-n$. Disclinations are topological defects of the
rotational order and have been subject to intense studies in the
context of graphene and
fullerenes.\cite{Gonzalez:1993,Lammert:2000,Lammert:2004,Roy:2010} In
the cut-and-glue construction, the integer $f>0$ ($f<0$) has the
meaning of counting the number of removed (added) $\pi/3$-wedges, see
Fig.~\ref{fig:disclination}(a) and (b). Note that for $f>0$, an
isolated disclination forms the tip of a nano-cone.\cite{Naess:2009}

The properties of an isolated disclination in the topological phase of
the Haldane model have recently been studied
theoretically.\cite{Ruegg:2013PRL} The main observation was that an
isolated defect in the Chern-insulator phase with Chern number $C$
acts as a source of a fictitious flux
\begin{equation}
\phi_f={\rm sign}(C)\frac{f}{4}\phi_0\mod{\phi_0},
\label{eq:phi_f}
\end{equation}
which pierces the defect core, where $\phi_0=h/e$ is the quantum of
flux. The quantized Hall conductivity $\sigma_{xy}=Ce^2/h$ implies that 
an isolated defect binds
a fractional charge given by
\begin{equation}
q_f=\sigma_{xy}\phi_f=e|C|\frac{f}{4}\mod{e}.
\label{eq:q_f}
\end{equation}
The defect states show up as single in-gap levels in the local
density-of-states with an energy which increases for increasing $f>0$,
see Fig.~\ref{fig:disclination}(c). It has been
argued\cite{Ruegg:2013PRL} that measuring such defect states would
provide an alternative probe of the topological phase, in analogy with
dislocation modes in weak or crystalline topological
insulators.\cite{Ran:2009,Ran:2010,Juricic:2012,Slager:2013}

Finally, let us clarify in which sense we use the expression ``fractional charges". 
We first recap that the subject of this paper is a non-interacting model on interesting 
but static lattice geometries. Therefore, unlike quasiparticle excitations of fractional 
quantum Hall liquids, the fractional charges in our system are not emergent dynamical 
excitations. Rather, they are bound to topological defects in a classical field 
(describing the lattice), which couples to the fermion system. In this respect, 
the fractional charges we observe at disclinations are more closely related to 
Majorana modes in vortices of topological superconductors\cite{Read:2000} 
or the quantum number fractionalization at domain walls in polyacetylene.\cite{Su:1979} 
Similar to the aforementioned examples, we find that the quantum mechanical 
wave function associated with the fractional charge is exponentially localized 
at the defect for an infinite system, justifying the used terminology. We mention also 
that in-gap states localized at point defects on
the hexagonal lattice have previously been studied in other
contexts.\cite{Mudry:2007, Ruegg:2011, He:2013} Furthermore,
disclinations also attracted attention in two-dimensional crystalline
topological superconductors where Majorana bound-states can be
realized.\cite{Teo:2013,Gopalakrishnan:2013}
\subsection{Implications from particle-hole symmetry}
\label{sec:ph}
In the presence of discrete
symmetries,\cite{Schnyder:2008,Kitaev:2009} a topological
classification of topological defects exists.\cite{Teo:2010} Here, we
focus on the role of particle-hole symmetry (class D) which gives rise
to a $\mathbb{Z}_2$ classification of point defects in two dimensions.
The $\mathbb{Z}_2$ index signals the presence or absence of a single
$E=0$ bound state. In the case of a superconductor, the $E=0$ mode
corresponds to a Majorana bound state while in a spin-polarized
insulator, the non-trivial defect binds a fractional charge $e/2$. As
long as the particle-hole symmetry is preserved, a trivial defect can
not be deformed into a non-trivial defect without closing of the bulk
gap. As we discuss in the following, the bound states of disclinations
in the Haldane model can be understood from this perspective.

We first discuss the condition for particle-hole symmetry in the
Haldane model which implies a specific form of the (first quantized)
Hamiltonian matrix $\hat{h}$. We write $\hat{h}$ in a sublattice basis
as
\begin{equation}
\hat{h}=\begin{pmatrix}
\hat{h}_{AA}&\hat{h}_{AB}\\
\hat{h}_{AB}^{\dag}&\hat{h}_{BB}
\end{pmatrix}
\label{eq:h}
\end{equation}
and denote the eigenfunction of $\hat{h}$ with energy $E$ as
$\psi(j)$:
\begin{equation}
\sum_j\hat{h}_{ij}\psi(j)=E\psi(i).
\end{equation}
A particle-hole symmetric spectrum is guaranteed if the particle-hole
conjugate wave function $\varphi(i)=\sigma_z\psi(i)^*$ is an
eigenstate of $\hat{h}$ with energy $-E$:
\begin{equation}
\sum_j\hat{h}_{ij}\varphi(j)=-E\varphi(i).
\label{eq:h_conjugate}
\end{equation}
Here, $\sigma_z$ is the third Pauli-matrix acting on the sublattice
degree of freedom (A-B). Equation~\eqref{eq:h_conjugate} implies that
\begin{equation}
\sigma_z\hat{h}^*\sigma_z=-\hat{h}.
\label{eq:ph}
 \end{equation}
or, using Eq.~\eqref{eq:h},
\begin{equation}
  \hat{h}_{AB}^*=\hat{h}_{AB},\quad 
  \hat{h}_{AA}^*=-\hat{h}_{AA},\quad
  \hat{h}_{BB}^*=-\hat{h}_{BB}.
\label{eq:hAB}
\end{equation}
In other words, a particle-hole symmetric spectrum is guaranteed if
the hopping between A and B sublattices is real but purely imaginary
among either A or B sites.\cite{Zheng:2011} Notice that the
particle-hole symmetry in the Haldane model relies on the
bipartiteness of the honeycomb lattice. Therefore, despite the formal
analogy, it is physically very distinct from the built-in
particle-hole symmetry of a superconductor in the Bogoliubov-de Gennes
description. In particular, lattice defects in the Haldane model have
the potential to violate the symmetry. In the following, we discuss
how this fact can be used to deduce certain properties of an isolated
disclination.
\begin{figure}
\includegraphics[width=1\linewidth]{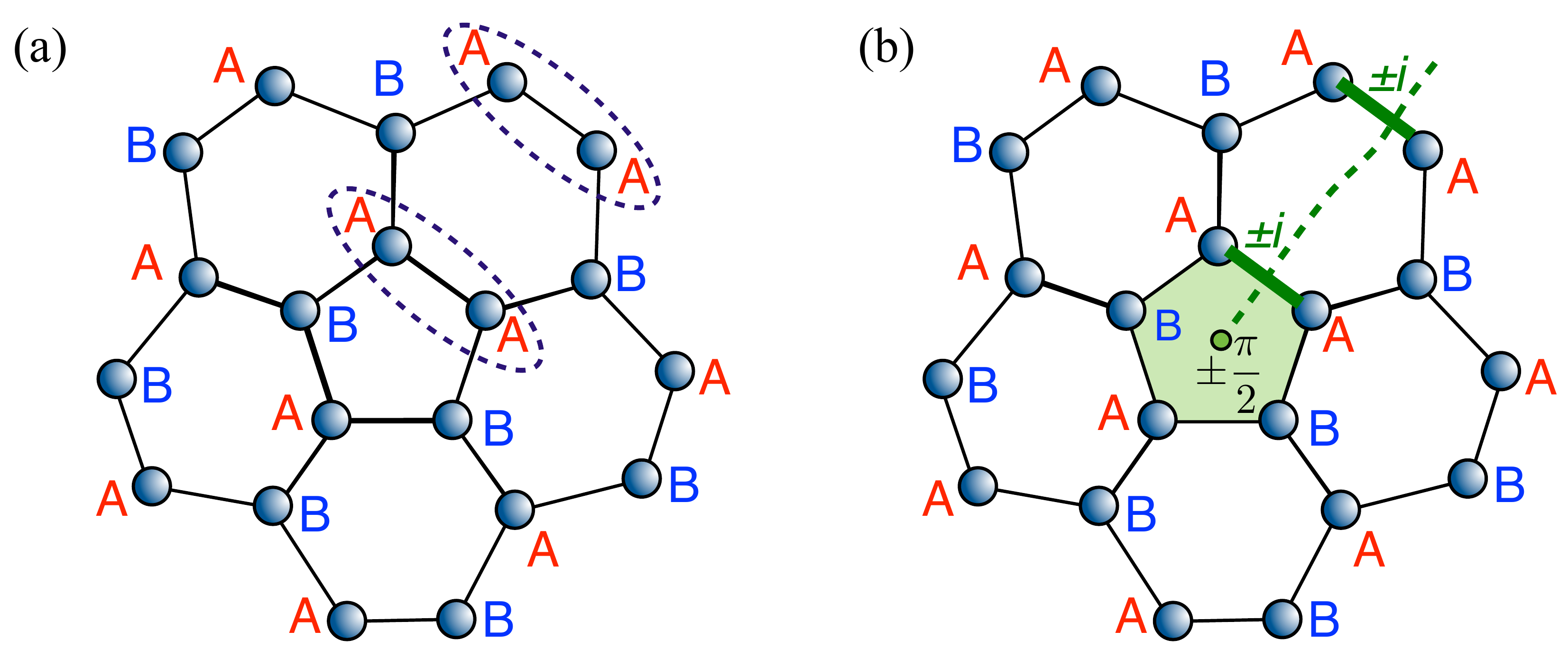}
\caption{(Color online.) (a) The $f=1$ disclination violates the
  particle-hole symmetry of the Haldane model with purely imaginary
  second neighbor hopping because two A sites meet across the seam.
  (b) Particle-hole symmetry can be restored by piercing the defect
  core with an external flux $\phi_e=\pm\phi_f/4$.}
\label{fig:restore_ph}
\end{figure}

For a disclination with odd $f$, the particle-hole symmetry is
violated. This is easy to understand because two A sites (or two B
sites) meet across the seam, as illustrated in
Fig.~\ref{fig:restore_ph}(a) for the case $f=1$. Hence, the conditions
\eqref{eq:hAB} are violated. However, one can restore the
particle-hole symmetry by piercing the center of the defect with an
external flux $\phi_e=\pm \phi_0/4$: if we bring the Dirac string into
line with the seam, as shown in Fig.~\ref{fig:restore_ph}(b), then all
the bonds crossing the Dirac string acquire an additional phase factor
$\pm i$. In particular, the nearest-neighbor hopping between two A
sites across the seam becomes purely imaginary. Similarly, the
second-neighbor hopping between A and B sites across the seam becomes
real. Thus, with an external flux $\phi_e=\pm\phi_0/4$, the conditions
\eqref{eq:hAB} are fulfilled and the spectrum is particle-hole
symmetric again. As a consequence, we know that the charge bound to
the defect is either 0 or $e/2 \mod{e}$:
\begin{equation}
q=q_f\pm\frac{eC}{4}=0 \mbox{ or } e/2\mod{e},
\label{eq:q}
\end{equation}
where $q_f$ is the intrinsic defect charge and $\pm eC/4$ is the
charge induced by the external flux. If in addition $C$ is odd, we
immediately conclude that $q_f=\pm e/4$. Hence, there is always a
non-trivial bound state. Using the linearity in $f$, we find that the
bound charge for a general $f$ is $q_f=\pm f e/4\mod{e}$. Thus, for
odd $C$, there is a $\mathbb{Z}_4$ classification of disclination
defects. For even $C$, Eq.~\eqref{eq:q} does not provide additional
information.

From the discussion above it is clear that the disclination bound
states are independent of a specific model as long as the
particle-hole symmetry is realized via the conditions~\eqref{eq:hAB}.
It then follows that for even $f$ the particle-hole symmetry is
preserved and the bound state (if present) is protected against any
local perturbation which preserves the conditions~\eqref{eq:hAB}.
Similarly, if $f$ is odd, the bound state (if present) is protected
against local perturbations which respect Eq.~\eqref{eq:hAB} in the
presence of an external flux $\phi_e=\pm\phi_0/4$. Hence,
particle-hole symmetry allows for a sharp topological distinction of
the defect states.

One may object that particle-hole symmetry in electronic systems is a
fine-tuned symmetry. In the Haldane model, it is for example easily
broken if the phases $\phi_{ij}$ of the second neighbor hopping are
tuned away from $\pm\pi/2$. Fortunately, direct diagonalization of the
tight-binding problem in the presence of particle-hole symmetry
breaking terms indicates that Eqs.~\eqref{eq:phi_f} and \eqref{eq:q_f}
still hold.\cite{Ruegg:2013PRL} This suggests that the results are
valid beyond the particle-hole symmetric limit. Note, however, that in
order to consistently define an electronic model with an invisible
seam in the presence of a disclination, it has to respect (at least on
average) the $C_3$ symmetry for even-$f$ and the $C_6$ symmetry for
odd-$f$ disclinations. The bound states are therefore only protected
in the presence of these crystalline symmetries and it is an
interesting open problem to show if the presence or absence of bound
states can be related to appropriate rotation
eigenvalues.\cite{Fang:2012,Teo:2013, Fang:2013} In the appendix, we
provide such a connection on the basis of the continuum description.

\section{Numerical results}
\label{sec:num}
\subsection{General considerations}
We now return to the study of the spherical nano surfaces and in the
following, we present the results obtained from numerically
diagonalizing Eq.~\eqref{eq:Haldane} on various geometries.  Because
of the finite size of the molecules, there are two important
differences to the case of an isolated disclination discussed in
Sec.~\ref{sec:disclination}.  First, apart from the energy scale of
the bulk gap $E_g~\sim |t_2|$, the finite size effect introduces an
additional scale given by the mean level-spacing $E_{R}\sim t/N_s$
where $N_s$ denotes the number of sites. We expect that only in the
regime $E_g\gg E_{R}$ it is possible to spot putative in-gap states
which are well separated from states of the quasi-continuum. Second,
there is always an even number of corner states and they generically
couple to each other, allowing in principle to push the in-gap states
into the quasi-continuum. In the regime $E_g\gg E_R$, the coupling is
expected to be weak and in-gap states are well-defined.
 
Section~\ref{sec:top} demonstrates that one can indeed identify in-gap
states for sufficiently large systems, consistent with the analysis of
isolated defects. By increasing the staggered sublattice potential in
the octahedral system, we also demonstrate that the in-gap states are
lost if the planar parent system is tuned into the trivial insulator
with $C=0$. In Sec.~\ref{sec:finite}, we discuss the finite size
effects. We therefore consider how the limit $E_g\gg E_R$ is
approached by tuning either the bulk gap via $|t_2|$ or the mean-level
spacing via the system size $N_s$.
\subsection{Corner states}
\label{sec:top}

\begin{figure}
\includegraphics[width=1\columnwidth]{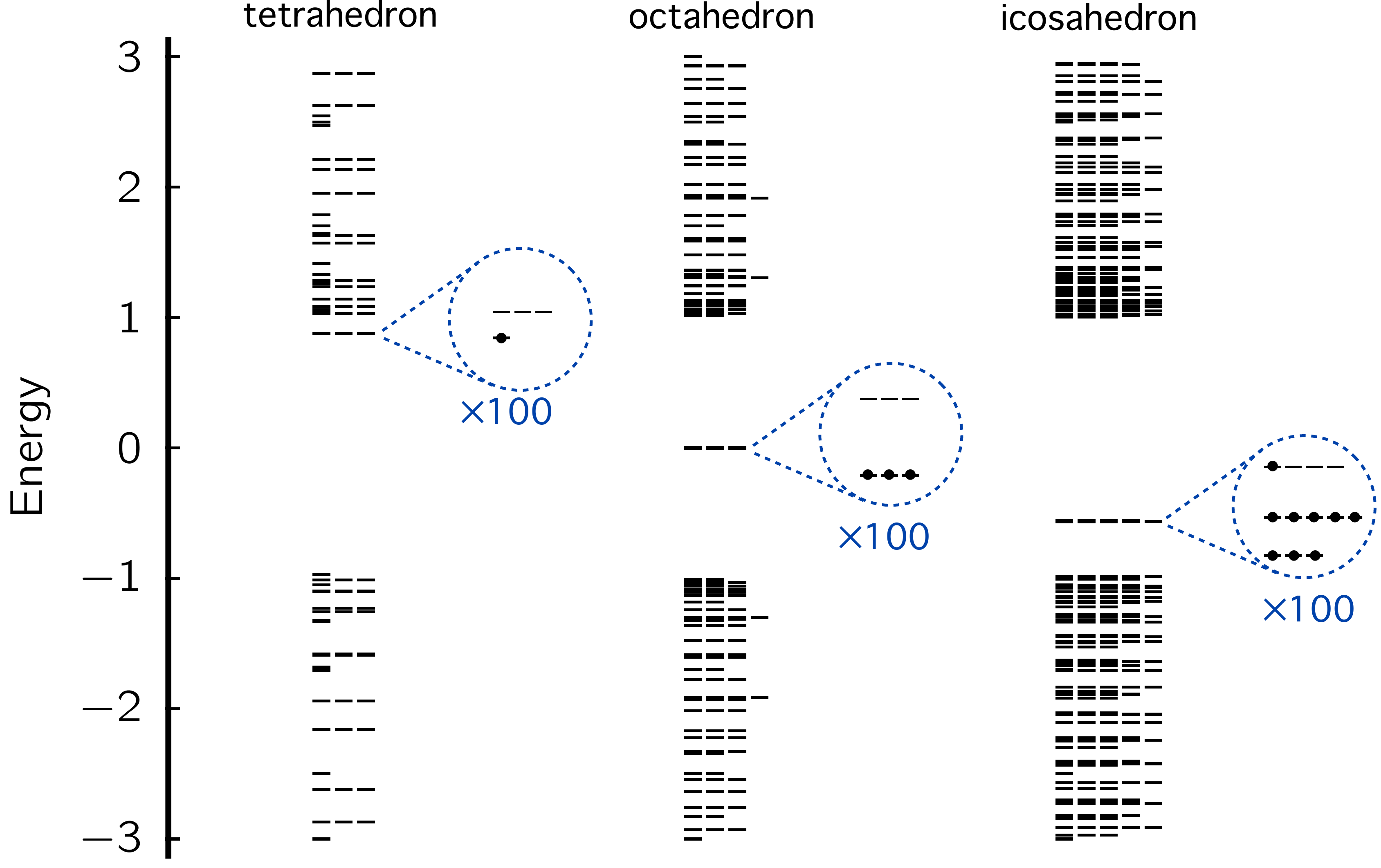}
\caption{Energy levels of tetrahedron, octahedron, and icosahedron
  model. Corner states are clearly visible in the gap and are
  separated from the remaining states. The inset shows a zoom-in (by a
  factor 100) of the in-gap states displaying the characteristic
  degeneracies 1+3 for tetrahedron, 3+3 for the octahedron, and 3+5+4
  for the icosahedron. In-gap levels which are occupied at half
  filling are marked with a dot. Parameters of the model Eq.~\eqref{eq:Haldane} are $t=1.0$,
  $|t_2|=0.2$, and $\phi=\pi/2$.}
\label{fig:spectrum}
\end{figure}

Figure~\ref{fig:spectrum} shows the spectrum for the tetrahedron, the
octahedron and the icosahedron model with parameters $|t_2|=0.2$,
$\phi=\pi/2$ and $\Delta=0$. For these parameters, the bulk system
realizes a Chern insulator with Chern number $C=\pm 1$.
The nano-surfaces considered in Fig.~\ref{fig:spectrum} have
$N_s=100$, $200$ and $500$ atoms for the tetrahedron, octahedron and
icosahedron, respectively. This choice guarantees that the distance
between the corners are roughly the same for the different geometries.
For all the systems, the condition $E_R\ll E_g$ is fulfilled. Indeed,
one can identify a quasi-continuum of states separated by a gap. In
addition, each spectrum features a characteristic number of in-gap
states (some of which are degenerate, as shown in the inset): $1+3=4$
for the tetrahedron, $3+3=6$ for the octahedron and $3+5+4=12$ for the
icosahedron. While the observed splitting of the in-gap level is
always the same and will be discussed later in Sec.~\ref{sec:eff}, the
order of the levels depends on details such as system size or the
ratio $t_2/t$.

For the given parameters, the spectrum of the octahedron model is
particle-hole symmetric as expected from the discussion in
Sec.~\ref{sec:ph}. On the other hand, particle-hole symmetry is
violated for the tetrahedron and icosaheron models.
The fractional charge bound to an isolated disclination can also be understood from
the spectra in Fig.~\ref{fig:spectrum} when considering the
half-filled systems for which the average charge per site is $e/2$. 
For the tetrahedron, one out of four in-gap states is filled. 
By symmetry, the wave function of this in-gap state has equal weight
on each of the four corners. Therefore, it contributes an average charge $e/4$
per corner. In the half-filled system, this charge has to compensate the fractional charge of the 
corner and we conclude that each defect carries a charge $-e/4$. Similarly, for the
octahedron, three out of six states are filled resulting in a charge
$-e/2$ per defect. Finally, for the icosahedron, nine out of twelve
levels are occupied resulting in $-3e/4$ per defect. These values are
in agreement with Eq.~\eqref{eq:q_f} obtained from the analysis of an
isolated disclination.

\begin{figure}
\includegraphics[width=1\columnwidth]{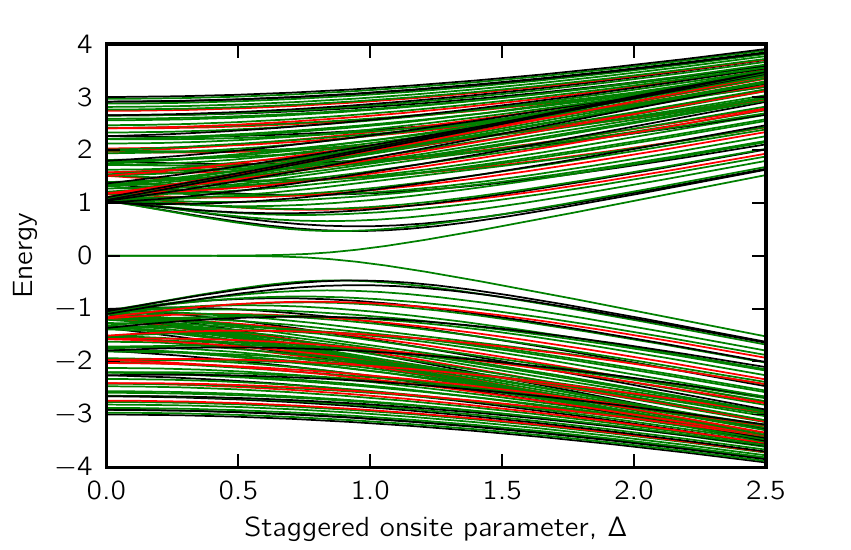}
\caption{(Color online.) The staggered sublattice potential $\Delta$
  on the octahedral nanosurface tunes a crossover between the
  topological (small-$\Delta$) and trivial (large-$\Delta$) regime.
  Parameters of the model Eq.~\eqref{eq:Haldane} with $N_s=392$ 
  sites are $t=1.0$, $|t_2|=0.2$, and $\phi=\pi/2$. Online version: Non-degenerate levels
  are colored black, 2-fold degenerate states red, 3-fold degenerate
  states green, and 4-fold degenerate states blue.}
\label{fig:Delta}
\end{figure}

The presence of non-trivial corner states is tied to the existence of a
non-trivial Chern number in the corresponding bulk system. This can
easily be tested by adding a staggered sublattice potential
Eq.~\eqref{eq:HDelta} which in bulk drives a transition to a gapped
phase with $C=0$. The corresponding result for an octahedral
nano-surface is shown in Fig.~\ref{fig:Delta}. As a function of the
sublattice potential $\Delta$, the spectrum changes
considerably. However, as opposed to the bulk system, finite size
effects prohibit a sharp closing of a gap between the small- and
large-$\Delta$ limit. Instead, a crossover at $\Delta\approx 1$ is
seen. Nevertheless, the small- and large-$\Delta$ regimes are clearly
distinct by the presence or absence of the corner states. Note that a
similar analysis for tetrahedral or icosahedral surfaces is not
possible because in an attempt to define a staggered sublattice
potential for these systems, the definition of A and B sites need to be
interchanged when crossing domain walls connecting two defects. These
domain walls can act as one-dimensional channels introducing
additional in-gap states in the large-$\Delta$
regime.\cite{Jung:2012,Zarenia:2012}

\subsection{Finite size effects}
\label{sec:finite}
The notion of in-gap states requires that $E_R\ll E_g$. If this
condition is not fulfilled, the corner states are no longer clearly
separated from the rest of the spectrum and a distinction between
topological and trivial regime (as demonstrated in
Fig.~\ref{fig:Delta}) is in general not possible. Despite this
expectation, we find that the finite-size effects on the octahedral
nano-surfaces have little consequences on the corner states making
them well-defined even if $E_R\sim E_g$. On the other hand, the corner
states of the tetrahedral and icosahedral surfaces are more sensitive
and indeed require $E_R\ll E_g$.
\begin{figure}
\includegraphics[width=1\linewidth]{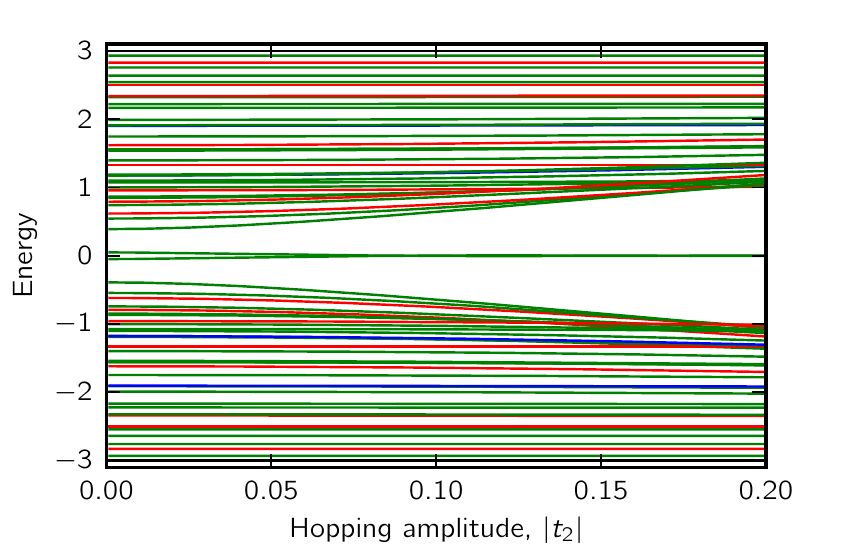}
\caption{(Color online.) Eigenvalues of the octahedron model as a
  function of the second neighbor hopping parameter amplitude $|t_2|$.
  Two 3-fold degenerate corner states occur in the gap and are clearly
  separated from other states when $|t_2|$ is about $0.05-0.1$. Phase
  of the second neighbor hopping parameter is $\phi=\pi/2$, the first
  neighbor hopping parameter equals 1.0 and $N_s=200$. Online version:
  Non-degenerate levels are colored black, 2-fold degenerate states
  red, 3-fold degenerate states green, and 4-fold degenerate states
  blue.}
\label{fig:t2_octahedron}
\end{figure}
\subsubsection{Octahedral nano-surfaces}
We start with the octahedral nano-surface. The spectrum as function of
$|t_2|$ for $N_s=200$ is shown in Figure~\ref{fig:t2_octahedron}. Note
that for the bulk system, $t_2=0$ corresponds to the gapless case
while a gap opens for non-zero $|t_2|$. On the other hand, the
spectrum of the octahedral nano-surface is discrete for any $t_2$.
Interestingly, the clearly separated corner states emerge out of a
pair of triplets near $E=0$ for small $t_2$ which already exists for
$t_2=0$. In this limit, these states are expected to be algebraically
localized at the corners\cite{Lammert:2000,Lammert:2004} while for
increasing $|t_2|$ the localization length decreases, making the
corner states increasingly better defined.

\begin{figure}
\includegraphics[width=0.9\linewidth]{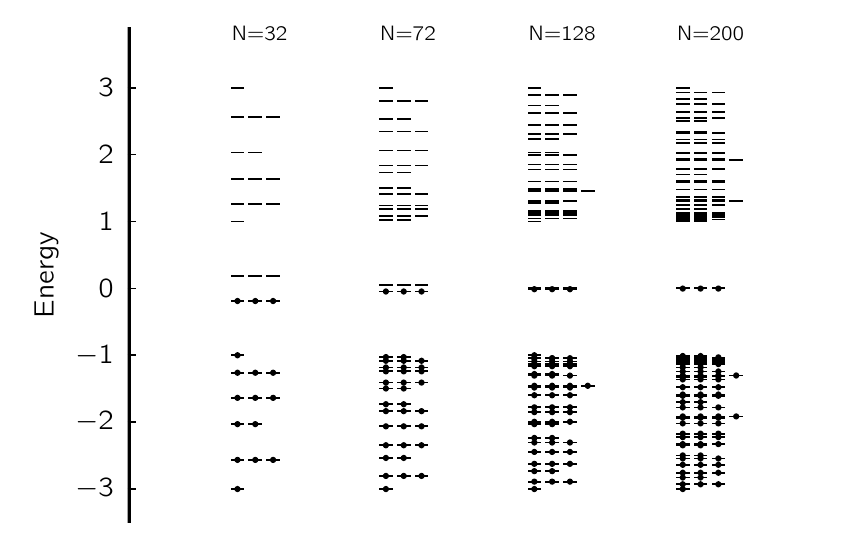}
\caption{Energy levels for octahedron as a function of system size (32
  sites to 200 sites). Occupied levels are indicated with a black dot,
  assuming that exactly half of all states (bulk and corner) are
  occupied. Parameters of the model are $t_1=1.0$, $|t_2|=0.2$, and
  $\phi=\pi/2$.}
\label{fig:Ns_oct}
\end{figure}

A similar trend is also observed for increasing system sizes at fixed
$|t_2|=0.2$, see Fig.~\ref{fig:Ns_oct}. The spectrum were obtained for
$N_s=32$, 72, 128 and 200 sites. For the larger systems with $N_s\geq
72$, corner states which are well separated from the quasi-continuum
are clearly visible. However, the characteristic pair of triplets
already exists for the smallest considered system with $N_s=36$ sites.

\subsubsection{Icosahedral nano-surfaces} 
\begin{figure}
\includegraphics[width=1\linewidth]{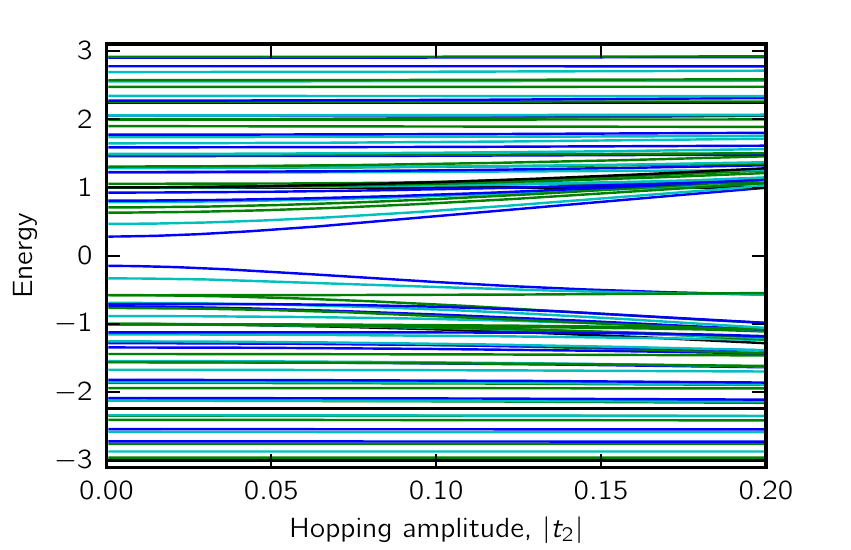}
\caption{(Color online.) Eigenvalues of the icosahedron model as a
  function of the second neighbor hopping parameter amplitude $|t_2|$.
  In-gap states close to the valence band edge appear when $|t_2|$ is
  about $0.15-0.2$. Phase of the second neighbor hopping parameter is
  $\phi=\pi/2$, $t=1$ and $N_s=320$. Online version: Non-degenerate
  levels are colored black, 2-fold degenerate states red, 3-fold
  degenerate states green, and 4-fold degenerate states blue.}
\label{fig:t2_icosahedron}
\end{figure}

\begin{figure}
\includegraphics[width=0.9\linewidth]{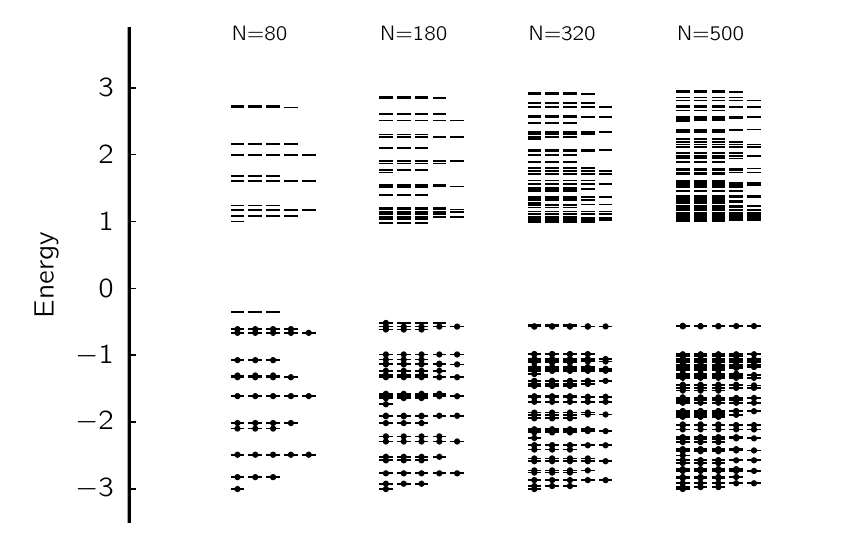}
\caption{Energy levels for the icosahedron as a function of system
  size (80 to 500 sites). Occupied levels are indicated with a black
  dot, assuming that exactly half of all states (bulk and corner) are
  occupied. Parameters of the model are $t_1=1.0$, $|t_2|=0.2$, and
  $\phi=\pi/2$.}
\label{fig:Ns_ico}
\end{figure}

We now turn to the icosahedral nano-surface.
Figure~\ref{fig:t2_icosahedron} shows the spectrum as function of the
second-neighbor hopping amplitude $|t_2|$ for $N_s=320$. For small
$|t_2|$, the characteristic level structure is not yet formed. Only
when $|t_2|$ is around $0.15-0.2$, in-gap states, which are clearly
separated from the quasi-continuum, emerge close to the valence band
edge. A similar finite size effect is also observed in the spectrum
for fixed $|t_2|=0.2$ but variable system size $N_s$, see
Fig.~\ref{fig:Ns_ico}. For the smallest size with $N_s=80$, the in-gap
levels are not yet well-separated from the rest of the states.
However, they emerge for larger systems.
\subsubsection{Tetrahedral nano-surfaces} 
\begin{figure}
\includegraphics[width=1\linewidth]{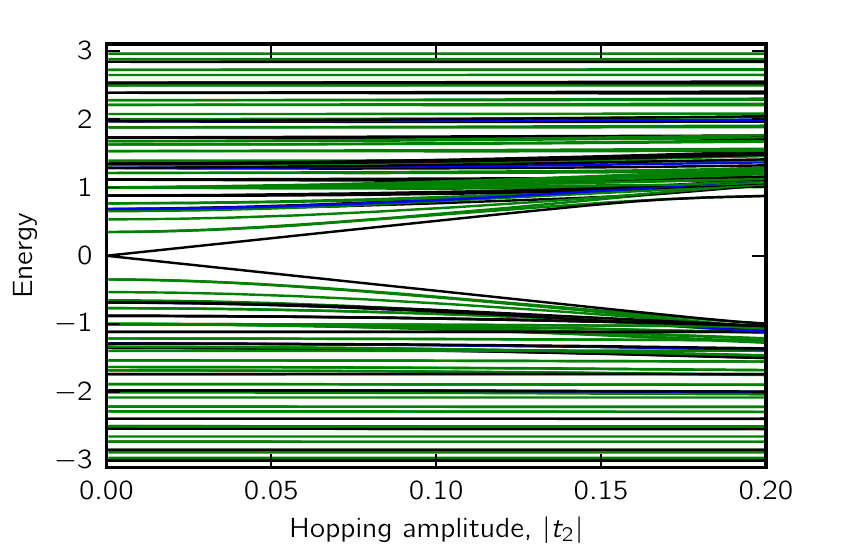}
\caption{(Color online.) Eigenvalues of the tetrahedron model as a
  function of the second neighbor hopping parameter amplitude $|t_2|$.
  In-gap states close to the conduction band edge appear for
  $|t_2|\approx 0.2$. Phase of the second neighbor hopping parameter
  is $\phi=\pi/2$, $t=1$ and $N_s=324$. Online version: Non-degenerate
  levels are colored black, 2-fold degenerate states red, 3-fold
  degenerate states green, and 4-fold degenerate states blue.}
\label{fig:t2_tetrahedron}
\end{figure}

\begin{figure}
\includegraphics[width=0.9\linewidth]{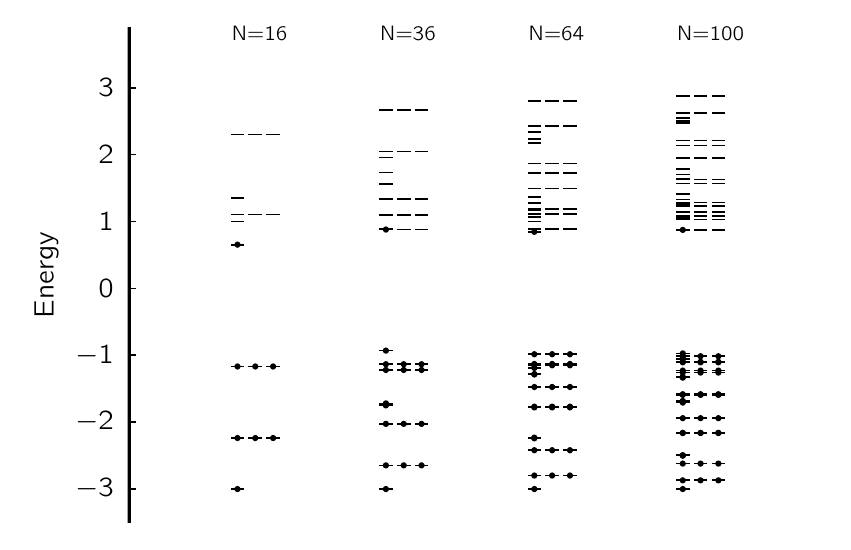}
\caption{Energy levels for the tetrahedron as a function of system
  size (16 to 100 sites). Occupied levels are indicated with a black
  dot, assuming that exactly half of all states (bulk and corner) are
  occupied. Parameters of the model are $t_1=1.0$, $|t_2|=0.2$, and
  $\phi=\pi/2$.}
\label{fig:Ns_tet}
\end{figure}

Eventually, we also discuss the finite size effects for the
tetrahedral systems where they appear to be strongest.
Figure~\ref{fig:t2_tetrahedron} shows the dependence of the spectrum
on $|t_2|$ for $N_s=324$. We find that a sizable second-neighbor
hopping amplitude of $|t_2|\sim 0.2$ is required to identify in-gap
levels appearing close to the conduction band. Figure~\ref{fig:Ns_tet}
shows the spectrum for various system sizes at fixed $|t_2|=0.2$. Only
for the largest system with $N_s=100$ the corner states are more or
less well separated from the quasi-continuum of the remaining states.
\section{Splitting of corner levels}
\label{sec:eff}
\subsection{Overview}
The tight-binding calculations presented in the previous section
(Sec.~\ref{sec:tb}) demonstrated that if the bulk Hamiltonian is in the
Chern insulator phase, the electronic spectra of sufficiently large
polyhedral nano-surfaces contain in-gap states which are clearly
separated from the quasi-continuum of the remaining
states. Furthermore, the number of in-gap states equals the number of
corners of the polyhedron. However, because of the coupling
between the corner states, the degeneracy is lifted in a
characteristic way, as summarized in Fig.~\ref{fig:polyhedra}. The
goal of the present section is to better understand this corner-level
splitting. As will be discussed in the following, the splitting can be
understood by assigning a fixed chirality to the corner states.

In Sec.~\ref{sec:model_corner}, we first study a general tight-binding
model for the corner-states alone. We show that in order to obtain an
energy spectrum which is consistent with the observed lifting of the
degeneracy, two magnetic monopoles have to be placed inside the
polyhedron.

In Sec.~\ref{sec:chiral}, we relate this observation to the fact that
the corner states are eigenstates of the $n$-fold rotation operator
about an axis piercing the defect core. The angular momentum of these
states is given by the Chern number $C$. We then argue that this leads
to a non-trivial Berry phase contribution which can be represented by
magnetic monopoles inside the polyhedra.

\subsection{Effective model for corner states}
\label{sec:model_corner}
\begin{figure}
\includegraphics[width=1\linewidth]{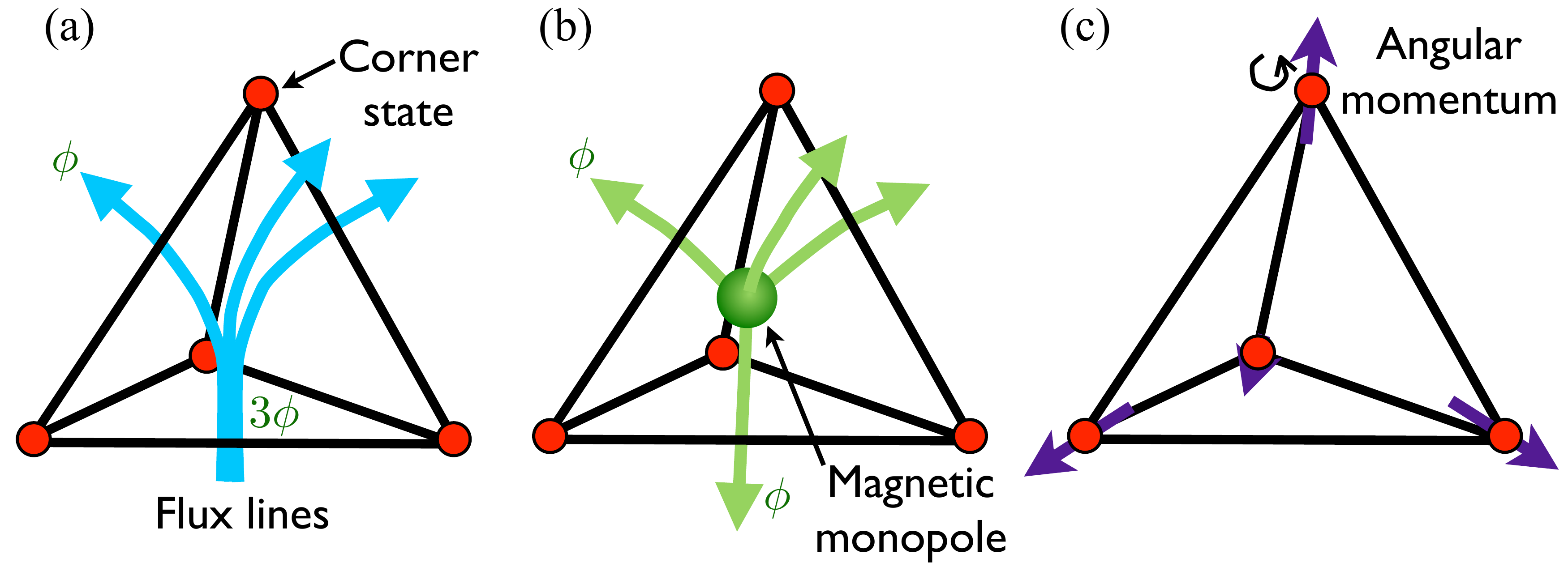}
\caption{(Color online.) Illustration of different corner-state models
  (here shown for the tetrahedron). (a) Tight-binding model describing
  hopping between corner states in the presence of an external
  magnetic flux. (b) If the fluxes through each triangle are equal
  modulo $\phi_0$, an equivalent representation with a magnetic
  monopole in the center of the polyhedron exists. (c) In the absence
  of external fluxes, electrons can pick up the same complex phase
  from a Berry phase term arising due to an internal angular
  momentum.}
\label{fig:eff_model}
\end{figure}

To study the splitting of the energy levels of the corner states, we
first introduce a phenomenological model. The model focuses on the
nearest-neighbor hopping processes between the corner states of the
different Platonic solids studied in this work:
\begin{equation}
  \mathcal{H}_{\rm corner}=\sum_{\langle i,j\rangle}
  \left(t_{ij}^{\rm eff}f_i^{\dag}f_j+{\rm h.c.}\right).
  \label{eq:Hcorner}
\end{equation}
Here, the sum runs over nearest-neighbor pairs and the operator
$f_i^{\dag}$ creates a corner state at the corner $i$. The hopping
amplitude between corner $i$ and $j$ is given by $t_{ij}^{\rm eff}$.
It turns out that in order to reproduce the observed level splitting,
it is crucial to allow for the possibility that the faces of the
polyhedra are threaded by a magnetic flux. Therefore, we assume
complex hopping amplitudes:
\begin{equation}
t_{ij}^{\rm eff}=|t_{\rm eff}|e^{ia_{ij}}.
\end{equation}
The total phase accumulated when hopping around a triangle with
corners $i$, $j$ and $k$ (labeled in a right-handed way) is then
related to the flux through the triangle $\phi_{ijk}$ by
\begin{equation}
a_{ij}+a_{jk}+a_{ki}=2\pi \frac{\phi_{ijk}}{\phi_0}
\end{equation}
where, as before, $\phi_0$ is the quantum of flux. By symmetry, we
expect that the flux through each triangle is identical. This requires
a configuration with an integer number of magnetic monopole quanta
inside the solid. To model this situation, we consider flux lines
which enter the solid through one face and then uniformly exit through
the remaining faces, as illustrated in Fig.~\ref{fig:eff_model}(a) for
the case of the tetrahedron.
Whenever the incoming flux is opposite equal to the outgoing flux
through one of the faces modulo $\phi_0$, such a flux line
configuration is indistinguishable from a magnetic monopole in the
center of the solid as shown in Fig.~\ref{fig:eff_model}(b). This
condition is only satisfied if the flux through a single face is given
by
\begin{equation}
\phi_{ijk}=n_{\Phi}\frac{\phi_0}{F}\mod\phi_0
\label{eq:Phi_F}
\end{equation}
where $n_{\Phi}$ is an integer and $F$ denotes the number of faces.
Equation~\eqref{eq:Phi_F} is just Dirac's quantization condition for
magnetic monopoles.

\begin{figure}
\includegraphics[width=0.9\linewidth]{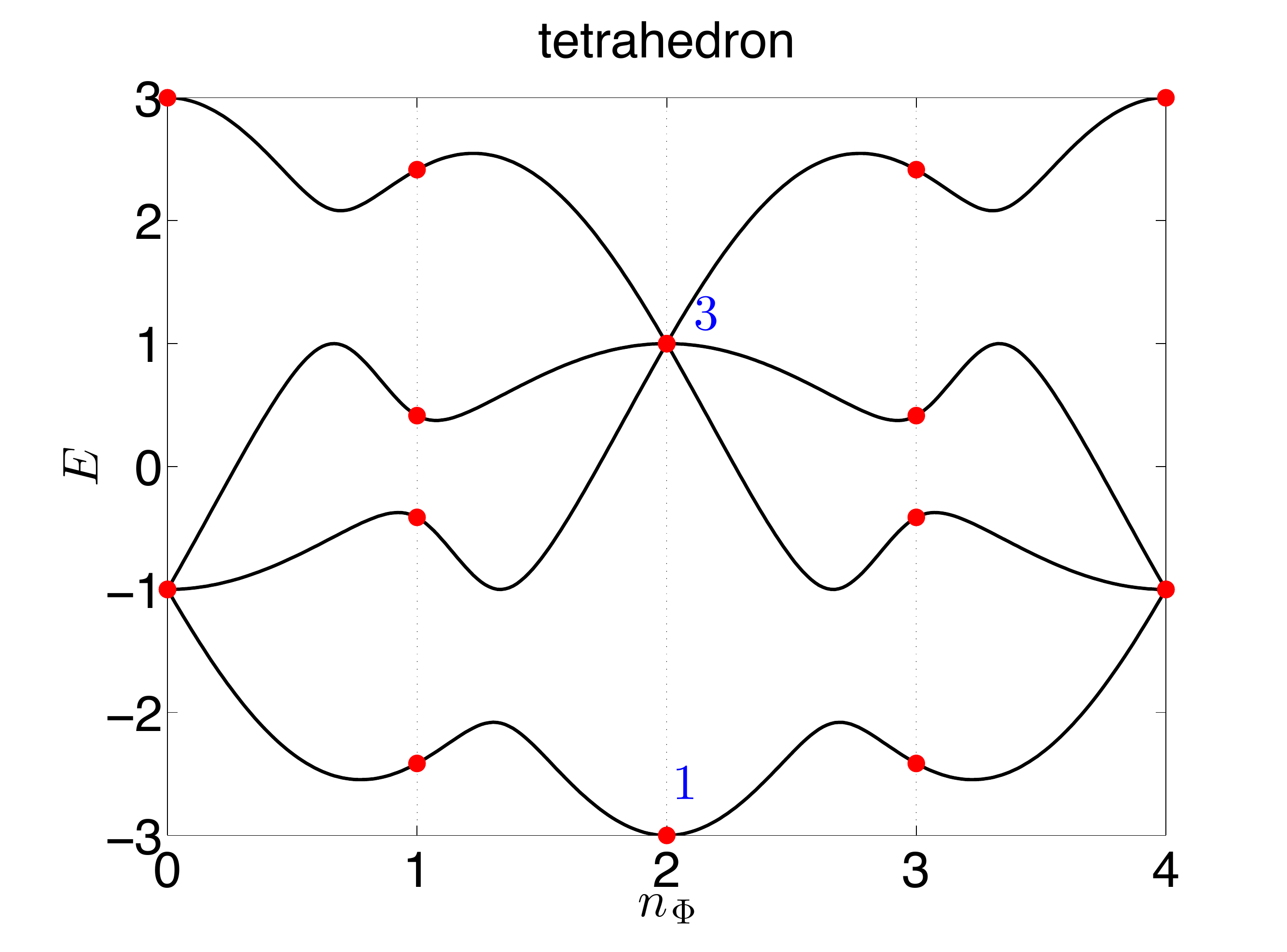}
\includegraphics[width=0.9\linewidth]{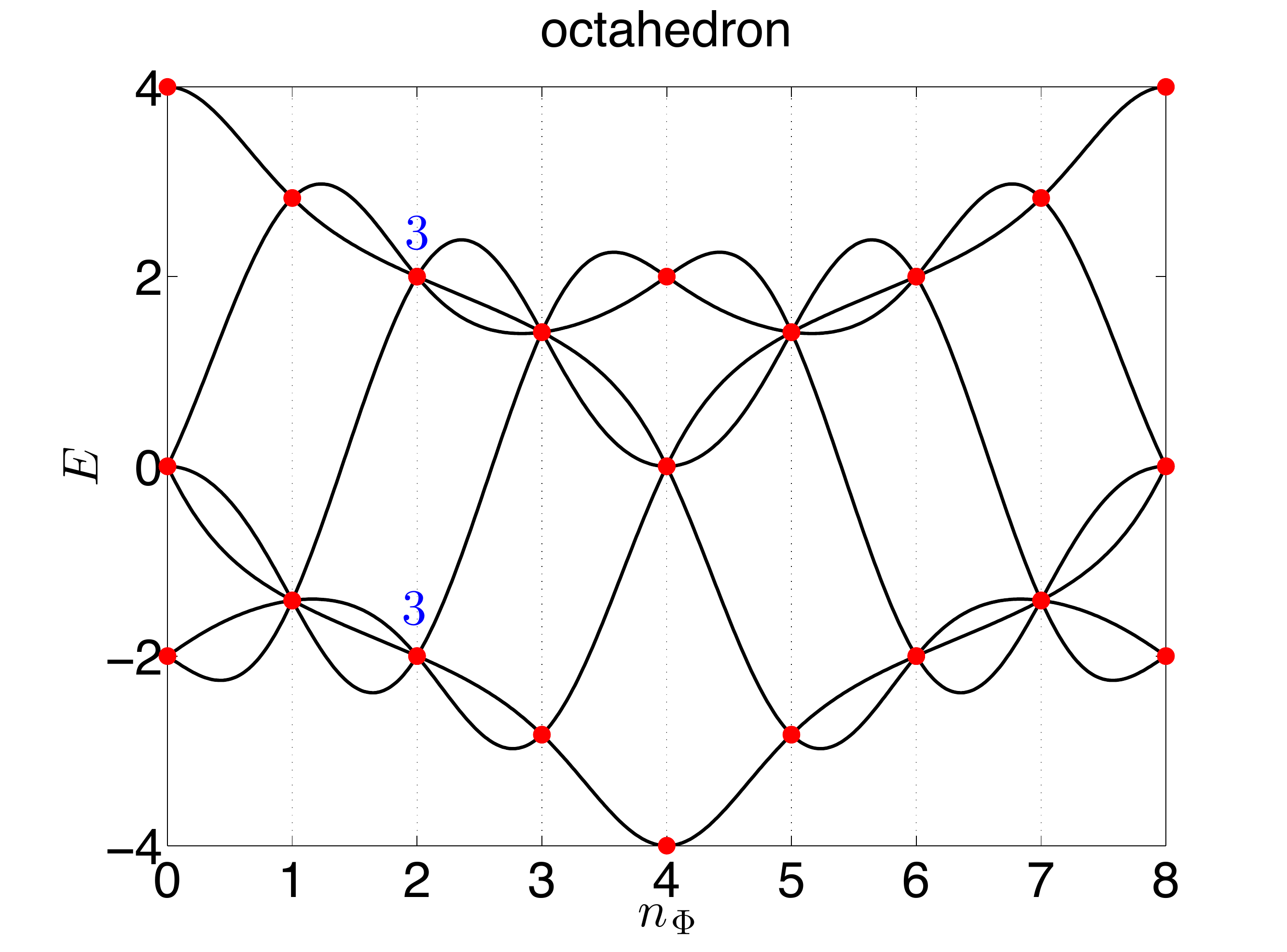}
\includegraphics[width=0.9\linewidth]{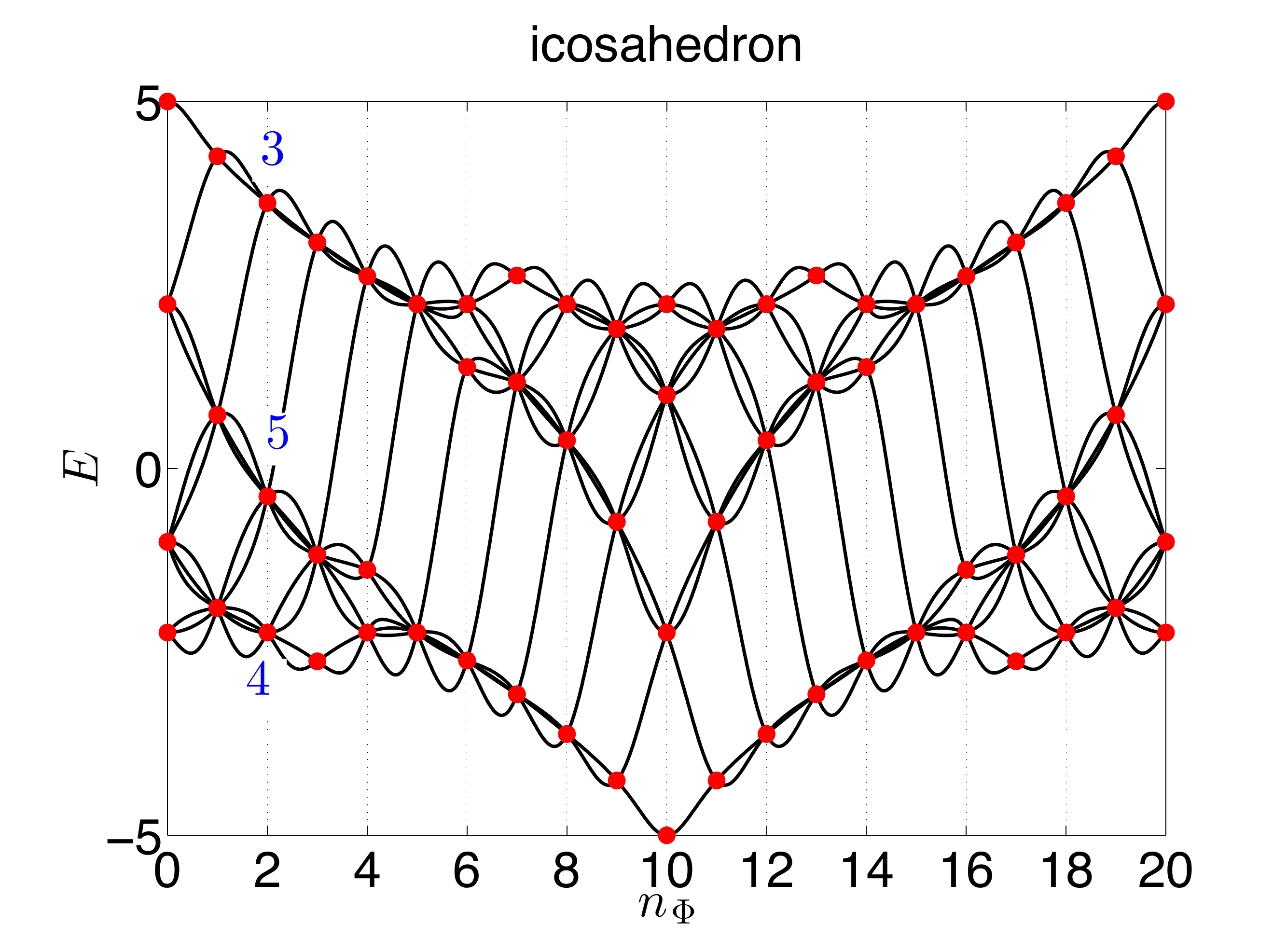}
\caption{The energy spectrum of the corner tight-binding model 
 on the tetrahedron, octahedron and icosahedron as function
  of the number $n_{\Phi}$ of inserted magnetic monopoles, see main
  text. The degeneracies for $n_{\Phi}=2$ are indicated in the plots.}
\label{fig:corner_hopping}
\end{figure}

The energy spectra of the model Eq.~\eqref{eq:Hcorner} as function of
the number $n_{\Phi}$ of elementary magnetic monopoles inserted into
the platonic solids are shown in Fig.~\ref{fig:corner_hopping}. As
described above, for non-integer values of $n_{\Phi}$, the flux
configuration does not correspond to Fig.~\ref{fig:eff_model}(b) but
to (a) with an outward pointing flux $\phi$ given by
Eq.~\eqref{eq:Phi_F}. Clearly, without a magnetic monopole
($n_{\Phi}=0$), the splitting and degeneracies are not consistent with
the numerical results of the full model shown in
Fig.~\ref{fig:spectrum}. Instead, a closer inspection shows that the
splitting for $n_{\Phi}=2$ is consistent for all the polyhedra,
i.e.~1+3 for the tetrahedron, 3+3 for the octahedron and 4+5+3 for the
icosahedron.

The $n_{\Phi}=2$ magnetic monopole, which has to be placed inside the
solids to reproduce the observed energy splitting, should not be
confused with the fictitious flux $\phi_f$ given in
Eq.~\eqref{eq:phi_f} (and which was also considered in the continuum
approximation to spherical
fullerenes\cite{Gonzalez:1992,Gonzalez:1993}). The fictitious flux
$\phi_f$ produces the corner states in the first place while the
$n_{\Phi}=2$ monopole is required to properly describe the coupling
between the corner states.

\subsection{Chiral corner states}
\label{sec:chiral}
What is the reason for the occurrence of the $n_{\Phi}=2$ monopole? In
the following, we argue that this is a result of the chiral nature of
the corner states. Indeed, the analysis of the continuum model for an
isolated disclination\cite{Ruegg:2013PRL} shows that the defect states
carry a finite angular momentum $J_z=C=\pm 1$ with respect to the
$n$-fold rotation axes through the center of the $n$-gonal defect. On
the polyhedral surfaces, the symmetry axis of the $C_n$ rotations
point outward through the corners of the polyhedron. Consequently,
when the electrons hop from corner to corner, the quantization axes
changes as well. As a result, if the electron hops around the triangle
with corners $i$, $j$ and $k$, it picks up a non-trivial Berry phase
given by
\begin{equation}
  \phi_{ijk}=\Omega({\bs e}_i,{\bs e}_j,{\bs e}_k)
  \sim{\bs e}_i\cdot({\bs e}_j\times{\bs e}_k)
\end{equation}
where $\Omega({\bs e}_i,{\bs e}_j,{\bs e}_k)$ is the solid angle
subtended by the three unit vectors pointing from the center of the
polyhedron to the three corners $i$, $j$ and $k$.  This Berry phase
can be represented by a magnetic monopole with $n_{\Phi}=2$, see
Fig.~\ref{fig:eff_model}(c). Note that similar Berry phase
contributions appear if an electron propagates in the background of
magnetic moments with non-coplanar order.\cite{Ohgushi:2000} We
present more details in the Appendix.

\section{Conclusions}
\label{sec:conclusion}
In summary, we have studied Haldane's honeycomb lattice model on
spherical nanosurfaces, namely the tetrahedron, the octahedron and the
icosahedron. For parameters which correspond to the Chern insulator
phase in the infinite planar model, we found that each corner of the
polyhedron carries a non-trivial bound state and we dubbed such
molecules {\it topological fullerenes}. In the energy spectrum, the
corner states show up as characteristic in-gap levels which are
clearly separated from the quasi-continuum of the remaining levels. We
related the occurrence of the corner states to the existence of
non-trivial defect states bound to isolated wedge-disclinations and
discussed the lifting of the degeneracies within an effective model
for the corner states. The presented example demonstrate that a
two-dimensional non-trivial bulk invariant can manifest itself in the
energy spectrum of a {\it closed} surface with no boundaries. While
our findings are based on the study of the Haldane model, we speculate
that similar results can be obtained in other models with odd Chern
number, such as the planar $p$-orbital model\cite{ZhangM:2011} or a
twisted version of Haldane's model.\cite{Alba:2011} We also expect
that our findings can be generalized to models with time-reversal
symmetry but non-trivial $\mathbb{Z}_2$ invariant, such as the
Kane-Mele model.\cite{Kane:2005a,Kane:2005b} In this case, the in-gap
modes would consist of Kramer's doublets and the bound states can
exhibit the phenomena of spin-charge
separation.\cite{Ran:2008,Qi:2008,Assaad:2013}

We now briefly comment on possible experimental realizations of
time-reversal invariant topological fullerenes. The first approach is
based on endohedral carbon fullerenes.\cite{Dunsch:2006} Following the
proposal to decorate graphene with 5d adatoms to induce a large
spin-orbit coupling,\cite{Hu:2012} we suggest that the icosahedron
model could be realized by instead enclosing 5d transition metal ions
within the sphere of the fullerenes. For the planar system, a
non-trivial $\mathbb{Z}_2$ invariant has been predicted (time-reversal
symmetry is preserved).\cite{Hu:2012} We therefore speculate that such
a non-trivial bulk $\mathbb{Z}_2$ invariant would give rise to
non-trivial corner states. Using the estimate
$\Delta_{SO}=3\sqrt{3}|t_2|=200$~meV for the spin-orbit induced
gap\cite{Hu:2012} and the value $t=2.7$~eV for the nearest-neighbor
hopping in graphene yields a ratio $t_2/t\approx 0.02$. Relating this
rough estimate to the findings of Sec.~\ref{sec:finite} shows that in
order to overcome the finite size effect, molecules should consist of
several hundred atoms. A second approach to topological fullerenes
would be the use of different materials with large intrinsic
spin-orbit coupling such as 2D bismuth or 2D tin. Owing to the buckled
nature of the honeycomb lattice realized in these
systems,\cite{Murakami:2006,Xu:2013} it is conceivable that these
materials would prefer to form octahedral nano-surface (for which it
is possible to globally define two sublattices). According to our
calculations, finite size effects are less pronounced for the
octahedral nano-surfaces and the corner states more likely to be
observed.

\acknowledgments A.R.\ acknowledges collaboration on related projects
with Chungwei Lin and Fernando de Juan and financial
support from the Swiss National Science Foundation. S.C.\ acknowledges 
discussion with David Vanderbilt and support by the Director, Office of Energy 
Research, Office of Basic Energy Sciences, Materials Sciences and Engineering 
Division, of the U.S. Department of Energy under contract DE-AC02-05CH11231 
which provided for the tight-binding calculations. J.E.M.\ acknowledges financial 
support from NSF DMR-206515.
\begin{appendix}
\section{Continuum description}
\subsection{Rotations and wedge disclinations}
\label{app:rot}

The (planar) Haldane model Eq.~\eqref{eq:Haldane} for $\Delta=0$
possesses a six-fold rotation symmetry about the center of a
hexagon. In the following, we review how this symmetry is implemented
in the effective low-energy description given by the following Dirac
Hamiltonian
\begin{equation}
H_{D}=-iv(\tau_z\sigma_x\partial_x+\sigma_y\partial_y)+m\tau_z\sigma_z.
\label{eq:HD}
\end{equation}
The Hamiltonian Eq.~\eqref{eq:HD} acts on a four component spinor
$\Psi=(\psi_A,\psi_B,\psi_{A'},\psi_{B'})$,
$\vec{\sigma}=(\sigma_x,\sigma_y,\sigma_z)$ are the Pauli matrices for
the sublattice (A-B) and $\vec{\tau}=(\tau_x,\tau_y,\tau_z)$ for the
valley ($K$-$K'$) degree of freedom. The mass term $m$ arises from a
finite $t_2$ in the topologically non-trivial phase.

Because the Dirac equation \eqref{eq:HD} is the low-energy
limit of a lattice model, spatial symmetries are realized differently
than for fundamental Dirac fermions.\cite{Mesaros:2010} Indeed,
translations and rotations need to account for the finite lattice
constant through the valley quantum number.  In particular, one can
identify two contributions to the rotation operator of physical
rotations by an angle $\alpha$ around the center of a
hexagon:\cite{Lammert:2004,Mesaros:2010}
\begin{equation}
R(\alpha)=R_{\rm lattice}(\alpha)R_{\rm Dirac}(\alpha).
\label{eq:R}
\end{equation}
Note that in the low-energy limit of Eq.~\eqref{eq:HD}, a continuous
rotation symmetry emerges
\begin{equation}
R(\alpha)^{\dag}H_{D}R(\alpha)=H_D
\end{equation}
with arbitrary $\alpha$. However, the restriction $\alpha=f\pi/3$ with
$f$ integer holds for physical rotations. Before providing the
explicit form of $R(\alpha)$ it is convenient to introduce a
symmetry-adapted basis with two new sets of Pauli
matrices\cite{deJuan:2013}
\begin{eqnarray}
\vec{\Sigma}&=&(\Sigma_x,\Sigma_y,\Sigma_z)=(\sigma_x\tau_z,\sigma_y,\sigma_z\tau_z)
\label{eq:Sigma}\\
\vec{\Lambda}&=&(\Lambda_x,\Lambda_y,\Lambda_z)=(\sigma_y\tau_x,\tau_z,-\sigma_y\tau_y)
\label{eq:Lambda}
\end{eqnarray}
$\vec{\Sigma}$ denotes the (pseudo) spin-1/2 degree of freedom arising
from the sublattice structure and $\vec{\Lambda}$ are the generators
of SU(2) rotations in valley space. In this basis, the Hamiltonian
Eq.~\eqref{eq:HD} in Fourier space is simply
\begin{equation}
H_D=v_F(\Sigma_x k_x+\Sigma_y k_y)+m\Sigma_z
\end{equation}
and $[\vec{\Lambda},H_D]=0$.

We now provide the explicit form of $R(\alpha)$. The first
contribution in Eq.~\eqref{eq:R} is the well-known rotation operator
for fundamental Dirac spinors
\begin{equation}
R_{\rm Dirac}(\alpha)=e^{i\frac{\alpha}{2}\left(\Sigma_z+2L_z\right)}
\end{equation}
where $L_z=-i(x\partial_y-y\partial_x)$ is the $z$-component of the
orbital angular momentum and $\Sigma_z=\sigma_z\tau_z$ the
$z$-component of the spin 1/2 degree of freedom (associated here with
the A-B sublattices). Hence, the generator for $R_{\rm Dirac}$ is the
sum of spin and orbital momentum. Note that $R_{\rm
  Dirac}(2\pi)\Psi=-\Psi$ which would make the wave-function double
valued when rotated by $2\pi$. The second contribution in
Eq.~\eqref{eq:R},
\begin{equation}
R_{\rm lattice}(\alpha)=e^{i\frac{3\alpha}{2}\Lambda_z},
\end{equation}
arises from the underlying lattice theory and compensates this minus
sign. Indeed, $R_{\rm lattice}(2\pi)\Psi=-\Psi$, so that the spinor is
single-valued under physical $2\pi$-rotations, $R(2\pi)\Psi=\Psi$. The
reason for the existence of $R_{\rm lattice}$ is the fact that the
Dirac cones are located at finite lattice momenta $K$ and $K'$. It
essentially accounts for the exchange of the valley and sublattice degrees
of freedom when a rotation by $\alpha=\pi/3$ is performed.

This analysis motivates to define the total angular momentum as
\begin{equation}
\vec{J}=\vec{L}+\frac{1}{2}\vec{\Sigma}+\frac{3}{2}\vec{\Lambda}.
\end{equation}
Because $[J_z,H_D]=0$, we can choose an eigenbasis of $H_D$ which
simultaneously diagonalizes $J_z$.\cite{Lammert:2004,Ruegg:2013PRL} 
In this basis, rotation by an angle $\alpha=f\pi/3$ acts as
\begin{equation}
R(f\pi/3)\Psi(r,\phi)=(\tau i)^f\Psi(r,\phi+f\pi/3).
\label{eq:RPsi}
\end{equation}
where $(r,\phi)$ are polar coordinates and $\tau=\pm 1$ denotes the
chirality of $\Psi$:
\begin{equation}
\Lambda_z\Psi(r,\phi)=\tau\Psi(r,\phi).
\end{equation} 
According to Eq.~\eqref{eq:RPsi}, for a wedge disclination, connecting
the wave function across the seam requires a non-trivial boundary
condition: the factor $(\tau i)^f$ precisely yields the fictitious
flux Eq.~\eqref{eq:phi_f}.
\subsection{Chiral defect states and Berry phase}
The solution of the continuum model in the presence of a
disclination\cite{Ruegg:2013PRL} shows that the bound state satisfies
\begin{equation}
\Lambda_z\Psi_0=-(2L_z+\Sigma_z)\Psi_0=\rm{sign}(C)\Psi_0.
\end{equation}
Thus, the defect states are eigenstates of $J_z$ with eigenvalues
$j_z={\rm sign}(C)$.

On a polyhedral surface, the quantization axis points outward through
the corners of the polyhedron. When hopping from corner to corner, the
quantization axis has to be adjusted which results in a non-trivial
Berry phase.  To calculate the Berry phase contribution, we note first
that for an isolated disclination, the azimuthal part of the
bound-state with $j_z=1$ is simply $e^{i \varphi}\sim
|p_x\rangle+i|p_y\rangle$. Next, we consider the surface of the sphere
and ask what is the overlap between two defect states which are
infinitesimally close to each other. We choose the spherical
coordinates such that the defect states have the same altitude
$\theta$ on the sphere but are separated along the ${\bs e}_{\phi}$
direction by an infinitesimal amount $\Delta \phi$. In spherical
coordinates, the first orbital is
\begin{equation}
|\psi^{(1)}\rangle=\frac{1}{\sqrt{2}}\left(
  |p_1(\theta,\phi)\rangle+i|p_2(\theta,\phi)\rangle\right)
\end{equation}
The second orbital is separated by $\Delta\phi$ in the direction ${\bs
  e}_{\phi}$ from the first orbital and is given by
\begin{eqnarray}
  |\psi^{(2)}\rangle&=&\frac{1}{\sqrt{2}}\left(
    |p_1(\theta,\phi+\Delta\phi)\rangle+i|p_2(\theta,\phi+\Delta\phi)\rangle\right)\\
  &=&(1-i\cos\theta\Delta\phi)|\psi^{(1)}\rangle
  -\frac{i}{\sqrt{2}}\sin\theta\Delta\phi|p_3(\theta,\phi)\rangle\nonumber
\end{eqnarray}
The effective hopping amplitude between the two states can now be
obtained from the overlap:
\begin{equation*}
t^{\rm eff}_{\Delta\phi}=-t\langle\psi^{(1)}|\psi^{(2)}\rangle= -t(1-i\cos\theta \Delta\phi)=-te^{ia_{\phi}\Delta\phi}
\label{eq:teff}
\end{equation*}
where the Berry connection is identified as
$a_{\phi}(\theta,\phi)=-\cos\theta$. Integrating along a closed path
from $\phi=0$ to $\phi=2\pi$ yields a Berry flux
\begin{equation}
\Phi_B=-2\pi\cos\theta \mod 2\pi.
\end{equation}
Hence, the Berry flux is identified with the solid angle enclosed by
the path of the electron on the sphere. For the hopping between the
corners of the polyhedron this result implies that each triangular
face is pierced by a flux
\begin{equation}
\Phi_F=4\pi/F.
\end{equation}
This is precisely Eq.~\eqref{eq:Phi_F} with $n_{\Phi}=2$ ($\hbar=1$).

\end{appendix}
\bibliography{biblio}

\end{document}